\documentclass[aps,prb,reprint,superscriptaddress,amsmath,floatfix]{revtex4-2}

\usepackage{graphicx}
\usepackage[colorlinks=true,linkcolor=black,citecolor=blue,urlcolor=blue]{hyperref}

\newcommand{\vc}[1]{\boldsymbol{#1}} 

\begin{document}

\title{
Orbital Order and Superconductivity in Bilayer Nickelate Compounds 
}

\author{Giniyat Khaliullin}
\affiliation{Max Planck Institute for Solid State Research,
Heisenbergstrasse 1, D-70569 Stuttgart, Germany}

\author{Ji\v{r}\'{\i} Chaloupka}
\affiliation{Department of Condensed Matter Physics, Faculty of Science,
Masaryk University, Kotl\'a\v{r}sk\'a 2, 61137 Brno, Czech Republic}

\begin{abstract}
We propose a theory for bilayer nickelate materials, where a large tetragonal
field -- intrinsic or induced by epitaxial strain -- lifts the orbital
degeneracy and localizes the \mbox{$3z^2\!-\!r^2$} orbital states. These
states host local spins $S=1/2$ bound into singlets by strong interlayer
coupling, and their dynamics is described by weakly dispersive singlet-triplet
excitations (``triplons''). The charge carriers occupy the wide bands of
\mbox{$x^2\!-\!y^2$} symmetry, and their Cooper pairing is mediated by the
high-energy triplon excitations. As the \mbox{$x^2\!-\!y^2$} band filling
increases, i.e., moving further away from the \mbox{Ni$^{3+}$} valence state,
the indirect Ruderman-Kittel-Kasuya-Yosida interactions between local spins
induce spin-density-wave order via triplon condensation. Implications of the
model for compressively strained \mbox{La$_3$Ni$_2$O$_7$} films and electron
doped oxychloride \mbox{Sr$_3$Ni$_2$O$_5$Cl$_2$} are discussed.
\end{abstract}

\date{\today}

\maketitle


Orbital degeneracy is a common feature of transition metal compounds and plays
a fundamental role in a wide range of physical phenomena, from magnetism to
metal-insulator and structural transitions~\cite{Ima98,Kho14,Kha05}. In recent
years, especially after the discovery of superconductivity (SC) in a number of
nickel-based oxides~\cite{Li19,Sun23,Wan24}, orbital physics has become a
central issue also in the field of high-temperature superconductivity. 

Nickelates are known for their rich spin, orbital, and electronic
properties~\cite{Med97,Ima98}. It is remarkable that the Ni valence states and
spin-orbital structures in the nickelate SCs vary broadly. While
\mbox{NdNiO$_2$}~\cite{Li19} with \mbox{Ni$^+$} ions and \mbox{$x^2\!-\!y^2$} 
orbitals fits well into the cuprate's ``single-layer, single-orbital, spin one-half'' 
paradigm, \mbox{La$_3$Ni$_2$O$_7$}~\cite{Sun23,Wan24} with mixed 
valence \mbox{Ni$^{2.5+}$} ions is a bilayer system, and 
both \mbox{$x^2\!-\!y^2$} and \mbox{$3z^2\!-\!r^2$} orbitals are essential. 

At ambient pressure, \mbox{La$_3$Ni$_2$O$_7$} shows magnetic order, and
becomes SC with $T_c\sim 80\:\mathrm{K}$ under high pressure. Theoretical
studies~(see, e.g., \cite{Sak24,Yan23b,Lia23,Yan23a,Liu23b,Lec23} and
references therein) suggest a strong pairing between the \mbox{$3z^2\!-\!r^2$}
electrons, while \mbox{$x^2\!-\!y^2$} band is mostly responsible for the SC
coherence within the planes. These theories are based on the multiorbital 
ab-initio band structures. The bilayer $t-J$ models
were also considered, suggesting that SC is driven by an effective interlayer
spin interaction between \mbox{$x^2\!-\!y^2$} orbitals~\cite{Oh23,Che24,Lu24}.
Experimentally, the key factors stabilizing SC in nickelates remain unclear, 
as the high pressure conditions restrict the use of standard probes; significant 
local inhomogeneity in the structural and electronic properties~\cite{LLiu25,Man25} 
adds further uncertainties.   

An important progress has been made recently: Under a compressive strain, the
bilayer nickelate films exhibit SC even at ambient
pressure~\cite{Ko25,Zho25,Liu25,Hao25,Hsu25,Osa25,LDnote}. Hole doping by oxidation or Sr
doping is also essential. These observations raise fundamental questions about
the microscopic origin of SC and its stabilization at ambient pressure,
especially about the role played by the lattice strain.

Motivated by these observations, we propose in this Letter a theoretical
model, which reveals a physical mechanism by which the in-plane compressive
strain stabilizes the superconductivity in bilayer nickelates. The general
phase behavior of the model, including magnetic ordering as a function of Ni
valence, is discussed. 

Our basic idea is that the compressive strain, which elongates the oxygen
octahedra, lifts the $e_g$ orbital degeneracy and localizes the
\mbox{$3z^2\!-\!r^2$} orbitals. The orbital ordering reduces the low-energy
Hilbert space available for the charge carriers, thus promoting
metal-insulator Mott transition. As an illustration, let us consider a single
layer nickelate with Ni \mbox{$d^7(t_{2g}^6e_g^1)$} configuration. The Mott
transition takes place when Coulomb repulsion between electrons on different
orbitals $U'=U-2J_\mathrm{H}$ ($J_\mathrm{H}$ is Hund's coupling) exceeds the
$e_g$ orbital bandwidth $W\sim 3-4\:\mathrm{eV}$. Due to $pd$-covalency,
effective interaction $U'$ in nickelates is rather small, comparable to $W$,
and they are typically metallic. In the presence of a tetragonal splitting
$\Delta$ of the $e_g$ level, however, the double occupation costs higher
energy, $U'+\Delta$, which may well exceed $W$ and thus stabilize the
\mbox{$3z^2\!-\!r^2$} orbital ordered Mott insulator. The case example is
nickel oxyhalides \mbox{Sr$_2$NiO$_3X$} ($X$=Cl, F), which are indeed
antiferromagnetic Mott insulators~\cite{Tsu22}, with an unpaired $e_g$
electron in the \mbox{$3z^2\!-\!r^2$} state.

\begin{figure}[tb]
\includegraphics[scale=1.0]{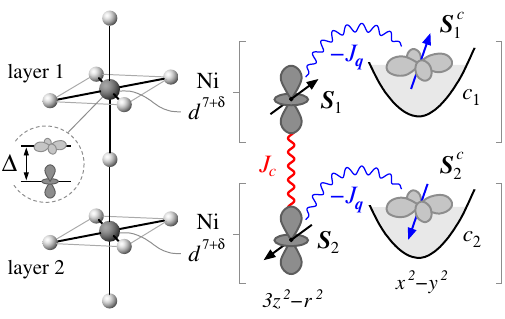}
\caption{
Schematic of the model for a bilayer nickelate, where large $e_g$ orbital
splitting $\Delta$ (see inset) stabilizes the \mbox{$3z^2\!-\!r^2$} states
hosting local spins $\vc S_1$ and $\vc S_2$. The strong interlayer exchange
coupling $J_c$ (red wavy line) binds $\vc S_1$ and $\vc S_2$ into a singlet
pair, forming a nonmagnetic background for a motion of the conduction
electrons in the $c_1$ and $c_2$ bands of \mbox{$x^2\!-\!y^2$} symmetry. The
itinerant spins $\vc S^c_1$ and $\vc S^c_2$ couple to local spins by
ferromagnetic interaction $-J_{\vc q}$ (blue wavy lines).
}\label{fig:model}
\end{figure}

{\it The model}.---The basic structure of our model is shown in
Fig.~\ref{fig:model}. The $e_g$ level splitting $\Delta\sim 1\:\mathrm{eV}$,
as evaluated in strained \mbox{La$_3$Ni$_2$O$_7$} films~\cite{Bhatta25,Yi25} and 
oxychlorides~\cite{Tsu22,Och25} is assumed to be large enough to localize an
electron in the \mbox{$3z^2\!-\!r^2$} states, which overlap mainly on the
interlayer $c$-axis bonds. The remaining $e_g$ electrons, with an average
density $\delta<1$ per Ni ion, occupy \mbox{$x^2\!-\!y^2$} states forming 
wide conduction bands. The local and itinerant electrons interact via the
Hund's and Kondo couplings. We now present the derivation of the model in
detail.

({\it i}) We start with the spin interactions between \mbox{$3z^2\!-\!r^2$}
orbitals. We use shorthand notations $x$ and $z$ for \mbox{$x^2\!-\!y^2$} and
\mbox{$3z^2\!-\!r^2$} states, correspondingly, and parameterize their
nearest-neighbor hoppings as follows. Strong overlap between $z$ orbitals on
the $c$-axis bonds gives the largest hopping denoted as $-t$, with $t>0$. The
orbital hoppings on the nearest-neighbor in-plane bonds are then given by
$t_{xx}=-\frac{3}{4}t$, $t_{zz}=-\frac{1}{4}t$, and $t_{xz}=\pm\frac{\sqrt
3}{4}t$ (upper/lower sign corresponds to $x/y$ bond direction)~\cite{SKnote}.

\begin{figure}[tb]
\includegraphics[scale=1.0]{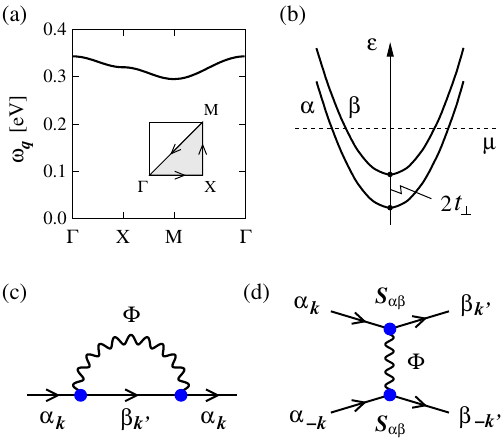}
\caption{
(a)~Momentum dependence of the singlet-triplet excitation energy 
$\omega_{\vc q}$, calculated with $t=0.6\:\mathrm{eV}$, $U=4.5\:\mathrm{eV}$, 
and $J_\mathrm{H}=0.6\:\mathrm{eV}$ along 
$\Gamma(0,0)$---$\mathrm{X}(\pi,0)$---$\mathrm{M}(\pi,\pi)$ path.
(b)~Schematic of the bonding $\alpha$ and antibonding $\beta$ band energies
$\varepsilon_{\vc k}$ split by interlayer hopping $t_{\perp}$~\cite{TBnote}. 
The chemical potential $\mu$ is determined by the \mbox{$x^2\!-\!y^2$} band 
filling $\delta$. 
(c)~Fermionic self-energy; the wavy line is the triplon propagator $\Phi$
describing singlet-triplet fluctuations, and the blue dots represent the spin
exchange coupling $J_{\vc k,\vc k'}$ between the local and itinerant
electrons.
(d)~Pairing interaction mediated by the singlet-triplet fluctuations.
}\label{fig:calc}
\end{figure}

Standard calculations~\cite{Kho14,Ole05} give the following expressions for
the interlayer $J_c$ and intralayer $J_{ab}$ exchange couplings between the
local spins $\vc S$ hosted by $z$ orbitals:
\begin{equation}\label{eq:Jc}
J_c= \frac{4t^2}{U}\,, \qquad 
J_{ab}/J_c=\frac{1}{16}\left(1-3\,\frac{J_\mathrm{H}}{U}\right). 
\end{equation}
The second term in $J_{ab}$ is due to the interorbital $t_{xz}$ hopping and
Hund's coupling in a virtual state. With $t\simeq 0.6\:\mathrm{eV}$ and
$U\simeq 4.5\:\mathrm{eV}$, appropriate for Ni ions, we evaluate $J_c\simeq
0.3\:\mathrm{eV}$. Because of the weak overlap $t_{zz}/t=-1/4$ of $z$ orbitals
within the planes, $J_{ab}$ is very small; a representative value of
$J_\mathrm{H}\simeq 0.6\:\mathrm{eV}$ gives $J_{ab}/J_c\simeq 0.04$. Thus the
local spins are tightly bound into interlayer singlet pairs of ${\vc S}_1$ and
${\vc S}_2$ (see Fig.~\ref{fig:model}). Following the bond-operator formalism
for antiferromagnets~\cite{Sac90,Som01}, we define triplon ${\vc T}$
operators: 
\begin{equation}\label{eq:T}
{\vc S}_{1,2}=\pm\frac{1}{2}\bigl({\vc T}+{\vc T}^\dagger\bigr) 
                -\frac{i}{2}\bigl({\vc T}^\dagger\times{\vc T}\bigr), 
\end{equation}
where upper/lower sign refers to the first/second layer. Triplons are
hard-core bosons describing singlet-triplet transitions within the $c$-axis
dimer bonds. The exchange interactions on these bonds
$J_c (\vc S_{i1} \cdot \vc S_{i2})$, and the in-plane interactions
$J_{ab} (\vc S_{i1} \cdot \vc S_{j1} + \vc S_{i2} \cdot \vc S_{j2})$
on the nearest-neighbor bonds $\langle ij \rangle$ are expressed via ${\vc T}$
operators: 
\begin{equation}\label{eq:HT}
H_T=\!\sum_{\vc q} A_{\vc q}{\vc T}_{\vc q}^\dagger{\vc T}^{\phantom{\dagger}}_{\vc q} +
\frac{1}{2} \sum_{\vc q} B_{\vc q}\left({\vc T}_{-\vc q}{\vc T}_{\vc q} +\mathrm{H.c.}\right) + \ldots , 
\end{equation}
where $A_{\vc q}=J_c+2J_{ab}\gamma_{\vc q}$, $B_{\vc q}=2J_{ab}\gamma_{\vc q}$, 
and $\gamma_{\vc q}=(\cos q_x +\cos q_y)/2$. Far from magnetic order,
triplon interactions, denoted by ellipsis in Eq.~\eqref{eq:HT}, and their hard-core nature 
can be neglected. The remaining quadratic form is diagonalized by Bogoliubov
transformation, leading to the triplon dispersion 
\begin{equation}\label{eq:wq}
\omega_{\vc q}=\sqrt{A_{\vc q}^2-B_{\vc q}^2} = J_c\sqrt{1+r\gamma_{\vc q}} \,, 
\end{equation}
where $r=4J_{ab}/J_c \ll 1$. As a result, local spin dynamics is described by
weakly dispersive singlet-triplet excitations at high energies 
$\omega_{\vc q}\simeq J_c\simeq 0.3\:\mathrm{eV}$ [see Fig.~\ref{fig:calc}(a)].

({\it ii}) Conduction electrons in the \mbox{$x^2\!-\!y^2$} orbitals form wide
bands, determined by large in-plane hopping $t_{xx}=-\frac{3}{4}t\sim
-0.45\:\mathrm{eV}$, like in cuprates. Finite interlayer hoppings of the
\mbox{$x^2\!-\!y^2$} orbitals are allowed due to their virtual hoppings
$t_{xz}$ to $3z^2-r^2$ orbital levels and hybridization with the oxygen $p$
states. These processes may operate within the $c$-axis rungs, 
$-t_\perp (c^\dagger_{i1} c_{i2} +\mathrm{H.c.})$, as well as connect the
farther neighbors, $-t'_\perp (c^\dagger_{i1} c_{j2}+\mathrm{H.c.})$ with 
$j \neq i$. The main effect of these terms is to form bonding
$\alpha_{\vc k}=\frac{1}{\sqrt 2}(c_1 + c_2)_{\vc k}$ and antibonding
$\beta_{\vc k}=\frac{1}{\sqrt 2}(c_1 - c_2)_{\vc k}$ states with the
dispersion relations
\begin{equation}\label{eq:xi}
\varepsilon_{\vc k}^\alpha= -3t\gamma_{\vc k} - t_\perp , \qquad
\varepsilon_{\vc k}^\beta = -3t\gamma_{\vc k} + t_\perp , 
\end{equation}
which are depicted in Fig.~\ref{fig:calc}(b). For simplicity, we omit here the
long-range hoppings, which are present in real materials and modify the Fermi
surface shapes~\cite{TBnote}, but not essential for our basic demonstration of
the model.

({\it iii}) Now we consider the interaction between local $\vc S$ and conduction
electron spins ${\vc S}^c=\frac12 c^\dagger_{s'}\hat{\vc \sigma}_{s's}c_s$.
There are two distinct spin exchange channels. First, the ferromagnetic Hund's
coupling $-2J_\mathrm{H}(\vc S \cdot \vc S^c)$. Second, the antiferromagnetic
Kondo interaction caused by virtual hoppings $t_{xz}$ between nearest-neighbor
$z_i$ and $x_j$ orbitals, $t_{xz}(x^\dagger_j z_i+z^\dagger_i x_j)$ with
$t_{xz}=\pm\frac{\sqrt 3}{4}t$. In $\vc k$-space, these processes lead to the
following term in the Hamiltonian:
\begin{equation}\label{eq:V}
\sum_{{\vc k}, \sigma} V_{\vc k} \sum_{i} \mathrm{e}^{i{\vc k}{\vc R}_i} 
\left(c^\dagger_{\vc k \sigma} z_{i\sigma} + \mathrm{H.c.}\right),
\end{equation}
where $V_{\vc k}=\frac{\sqrt 3}{2}t\,\eta_{\vc k}$ and $\eta_{\vc k}=(\cos k_x
-\cos k_y)$. Treating this term perturbatively, we obtain Kondo interaction:
\begin{equation}\label{eq:Kondo}
H_\mathrm{K}= 2 \sum_{\vc k, \vc k'} V_{\vc k} V_{\vc k'} 
\left(\frac{1}{E^+} + \frac{1}{E^-}\right) 
\left(\vc S_{\vc q} \cdot \vc S^c_{\vc k', \vc k}\right),
\end{equation}
where $\vc S^c_{\vc k', \vc k}=\frac12 c^\dagger_{\vc k' s'}
\hat{\vc \sigma}_{s's}c_{\vc k s}$, and $\vc k'=\vc k \!+\!\vc q$. 
$E^+$ ($E^-$) is the energy cost required to add (remove) an electron to
(from) a singly-occupied $z$ orbital level.

The total interaction between the local and itinerant spins within a bilayer
reads then as:
\begin{equation}\label{eq:Hint}
H_\mathrm{int}= - \sum_{\vc k, \vc k'} J_{\vc k, \vc k'} 
\left[ \left(\vc S_{\vc q} \cdot \vc S^c_{\vc k', \vc k}\right)_1 + 
       \left(\vc S_{\vc q} \cdot \vc S^c_{\vc k', \vc k}\right)_2 \right],
\end{equation}
where the coupling constant
\begin{equation}\label{eq:Jq}
J_{\vc k, \vc k'}\!= 
2J_\mathrm{H}\left(1-\kappa\,\eta_{\vc k}\eta_{\vc k'}\right)
\end{equation}
includes both Hund's and Kondo couplings. The latter brings about the momentum
dependence in the exchange process, and its relative strength is given by
\begin{equation}\label{eq:kappa}
\kappa= \frac{3}{4} \frac{t^2}{J_\mathrm{H} \widetilde E}\,, 
\quad\text{where}\quad
\frac{1}{\widetilde E}=\left(\frac{1}{E^+} + \frac{1}{E^-}\right). 
\end{equation}
The interorbital excitation energies $E^+$ and $E^-$ depend on various
parameters such as $U$, $J_\mathrm{H}$, intersite Coulomb interactions,
tetragonal splitting, doping dependent Fermi level position, etc. Using a
representative value of $\widetilde E \simeq 1\:\mathrm{eV}$, and 
$t \simeq J_\mathrm{H} \simeq 0.6\:\mathrm{eV}$, we arrive at a rough estimate
of $\kappa \simeq 0.5$. This implies that the overall coupling 
$J_{\vc k,\vc k'}$ is ferromagnetic, and thus the model is free of Kondo problem.
We also note that, in contrast to manganites with large total spin $S=2$, the
Hund's splitting in nickelates is smaller than the $e_g$ orbital bandwidth
$W\sim 3-4\:\mathrm{eV}$, i.e., well below the double-exchange limit of
$2J_\mathrm{H}>W$. For these reasons, we can safely treat the $J_{\vc k,\vc
k'}$ coupling \eqref{eq:Jq} effects perturbatively.  This is in contrast to
the case of effective models (e.g. \cite{Oh23,Che24,Lu24}) derived from a
strong Hund's coupling limit of $J_\mathrm{H} \rightarrow \infty$.

{\it Spin-singlet phase and superconductivity}.---We first rewrite
Eq.~\eqref{eq:Hint} in terms of a triplon field defined as
$\phi_\nu = \frac{1}{\sqrt 2} (T + T^\dagger)_\nu \equiv \frac{1}{\sqrt 2}
(S_1-S_2)_\nu$, where $\nu=(x,y,z)$. As the triplon density in the singlet
phase is small, we keep $\phi$-linear terms only: 
\begin{equation}\label{eq:Hintphi}
H_\mathrm{int}= - \frac{1}{\sqrt 2} \sum_{\vc k, \vc k'} 
J_{\vc k, \vc k'} \; \vc \phi_{\vc q} \cdot 
\left(\vc S_{\alpha \beta} + \vc S_{\beta \alpha}\right)_{\vc k', \vc k}\;.
\end{equation}
Here, spin density operator $(\vc S_{\alpha \beta})_{\vc k', \vc k} =
\frac12\alpha^\dagger_{\vc k' s'}\hat{\vc \sigma}_{s's}\beta_{\vc k s}$
connects bonding $\alpha$ and antibonding $\beta$ fermions, as dictated by the
odd parity of the triplon field. 

Without magnetic order, the triplon Green's function matrix is diagonal and
isotropic: $\Phi_{\nu\nu'} = \langle T_\tau \phi_\nu \phi_{\nu'} \rangle 
= \delta_{\nu\nu'} \Phi$. In harmonic approximation, i.e., neglecting triplon 
interactions, the bare Green's function is given by
\begin{equation}\label{eq:Phi0}
\Phi_0= \frac{J_c}{\omega^2 + J_c \Lambda_{\vc q}} \quad
\text{with}\quad \Lambda_{\vc q}=J_c(1+r\gamma_{\vc q}).
\end{equation}
This propagator describes singlet-triplet excitations with the energy
$\omega_{\vc q} = (J_c \Lambda_{\vc q})^{1/2}$ of Eq.~\eqref{eq:wq}. We use it
to evaluate the fermionic self-energy in Fig.~\ref{fig:calc}(c), and obtain
the mass enhancement factor of $(1+\lambda)$, where 
\begin{equation}\label{eq:lambda}
\lambda \approx \frac{3}{2}\frac{J_\mathrm{H}^2}{J_c} \;N(0) 
\end{equation}
quantifies the coupling constant in the model. $N(0)$ is the density of states
(per spin) at the Fermi level. We notice that $\lambda$ has the same structure
as in electron-phonon theories, with $J_\mathrm{H}$ and $J_c$ playing the
roles of electron-lattice coupling and phonon energy, respectively. As a rough
estimate for $\lambda$, we may take $N(0)\simeq 0.2-0.3\:\mathrm{eV}^{-1}$,
$J_\mathrm{H}\simeq 0.6\:\mathrm{eV}$, $J_c\simeq 0.3\:\mathrm{eV}$, and
obtain $\lambda \simeq 0.4-0.5$.

As in the electron-phonon problem, the same parameter $\lambda$ quantifies the
strength of the pairing interaction mediated by singlet-triplet excitations.
Figure~\ref{fig:calc}(d) shows that these excitations lead to the pair
correlations between the bonding $\alpha$ and antibonding $\beta$ states. We
find that the SC order parameters on these bands have an opposite sign,
$\Delta^\mathrm{SC}_\alpha = - \Delta^\mathrm{SC}_\beta$. This so-called
$s_{\pm}$ symmetry is common to spin-mediated pairing models, including those
proposed for nickelates~\cite{Yan23b,Liu23b}. However, the nature of the
pairing glue in our theory, i.e., high-energy triplon excitations provided by
the \mbox{$3z^2\!-\!r^2$} orbital sector, is radically different from the
previous models. Observing the triplons would be the ``smoking gun'' for our
theory.  

BCS mean-field treatment of the interaction in Fig.~\ref{fig:calc}(d) leads to
high values of $T_c\!\sim\!\Omega\exp(-1/\lambda)$, where $\Omega=J_c$ (or
Fermi energy $\varepsilon_F$ at small doping $\delta$), due to high excitation
energy $J_c\sim 0.3\:\mathrm{eV}$ and sizable coupling constant $\lambda$.
Even though a more quantitative treatment of the model would reduce this
estimate, it shows the great potential for high-$T_c$ SC in the bilayer
nickelates under a large tetragonal field.

\begin{figure}[tb]
\includegraphics[scale=1.0]{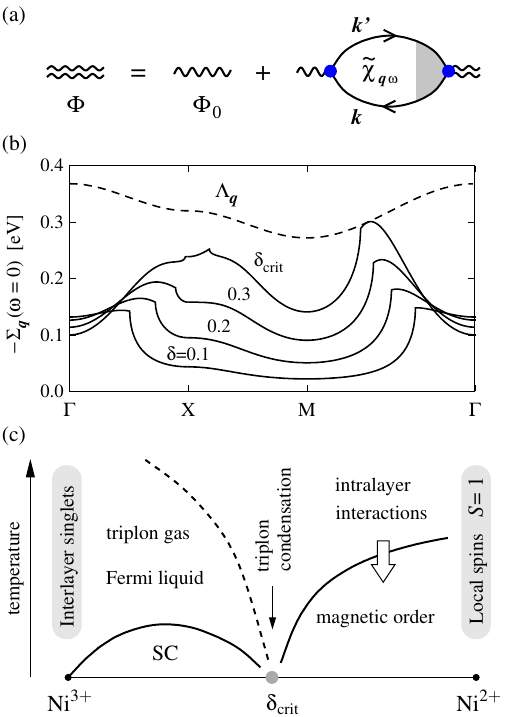}
\caption{
(a)~Triplon propagator $\Phi$ including the coupling $J_{\vc k,\vc k'}$ (blue
dots) between the local and itinerant spins. The self-energy depends on the
\mbox{$x^2\!-\!y^2$} band spin susceptibility $\widetilde\chi_{\vc q,\omega}$,
enhanced by Hubbard correlations (shaded vertex) between conduction electrons.
(b)~The triplon self-energy at $\omega=0$ for different dopings $\delta$ away
from the \mbox{Ni$^{3+}$} valence state. At $\delta_\mathrm{crit}\simeq 0.4$,
the self-energy reaches the dashed curve $\Lambda_{\vc q}$ at some $\vc q$
between $\Gamma$ and $\mathrm{M}$ points, and triplons condense into SDW
order. Parameters used: $J_\mathrm{H}=0.6\:\mathrm{eV}$, $\kappa=0.5$, and
$t=0.6\:\mathrm{eV}$. 
(c)~Phase behavior of the \mbox{$3z^2\!-\!r^2$} orbital-ordered bilayer
nickelate as a function of Ni valence. Close to the \mbox{Ni$^{3+}$} end, the
\mbox{$3z^2\!-\!r^2$} local spins are bound into the interlayer singlet pairs,
while the \mbox{$x^2\!-\!y^2$} band electrons form a Fermi-liquid;
singlet-triplet excitations (triplons) mediate their Cooper pairing. At large
doping, $\delta\!>\!\delta_\mathrm{crit}$, the intralayer RKKY interactions
between local spins, mediated by the band electrons, induce magnetic order via
triplon condensation. 
}\label{fig:PD}
\end{figure}

{\it Magnetic instability at large doping}.---We now consider the effects of
the exchange coupling $J_{\vc k, \vc k'}$ on local spins. Including this
coupling, we obtain the Green's function $\Phi$ in the following form:
\begin{equation}\label{eq:Phi}
\Phi= \frac{J_c}{\omega^2 + J_c (\Lambda_{\vc q} + \Sigma_{\vc q, \omega})}\,, 
\end{equation}
where the triplon self-energy
$\Sigma_{\vc q, \omega}= -4J_\mathrm{H}^2\,\widetilde\chi_{\vc q, \omega}$
is determined by the \mbox{$x^2\!-\!y^2$} band spin susceptibility
$\widetilde\chi_{\vc q, \omega}$, see Fig.~\ref{fig:PD}(a). Magnetic order,
i.e., triplon condensation, takes place when the static ($\omega=0$)
self-energy obeys the equation $\Lambda_{\vc q}+\Sigma_{\vc q}=0$. At a small
density $\delta$ of the band electrons, their mutual interactions can be
neglected, and $\Sigma_{\vc q}$ is given by the bare susceptibility 
$\chi_{\vc q}$, modified by the nonlocal nature of the exchange constant
$J_{\vc k,\vc k'}$ \eqref{eq:Jq}: 
\begin{equation}\label{eq:Sigma}
\Sigma_{\vc q} = -2J_\mathrm{H}^2 \sum_{\vc k} 
 \;(1-\kappa\,\eta_{\vc k}\eta_{\vc k'})^2 \;\,
 \frac{n_{\vc k}-n_{\vc k'}}{\varepsilon_{\vc k'}-\varepsilon_{\vc k}} \;.
\end{equation}
The self-energy evolves quickly with doping, especially at momenta away from
$\Gamma$ point [see Fig.~\ref{fig:PD}(b)], due to the Fermi surface shape and
$\kappa$-term effects. As a result, the magnetic instability criteria are
fulfilled already at $\delta\simeq 0.4$, i.e., well below the Mott limit for
\mbox{$x^2\!-\!y^2$} orbital sector. Physically, the magnetic order is driven
by the indirect exchange interactions between local spins $\vc S$, mediated by
the \mbox{$x^2\!-\!y^2$} band electrons. Indeed, the triplon self-energy [a
fermionic bubble in Fig.~\ref{fig:PD}(a) and Eq.~\eqref{eq:Sigma}] is nothing
but the Ruderman-Kittel-Kasuya-Yosida (RKKY) interaction between local spins,
and magnetic instability takes place when this interaction is large enough to
overcome singlet-triplet gap $\Lambda_{\vc q} \sim J_c$. There is a certain
analogy with the Doniach's RKKY-vs-Kondo competition in heavy-fermion 
systems~\cite{Don77,*[{see also Chapter 17 in }][{}]Col15}, though the origin
of a nonmagnetic phase here (local spin dimers) is much different from Kondo
singlet formation.

At the critical density $\delta_\mathrm{crit}$, a spin-density-wave (SDW)
order sets in. Approaching half-filling of the \mbox{$x^2\!-\!y^2$} band,
$\delta=1$, the enhanced Hubbard correlations are expected to cooperate with
Hund's coupling, resulting in an antiferromagnetic Mott insulator with
\mbox{Ni$^{2+}$} spins $S=1$. 

We sketch in Fig.~\ref{fig:PD}(c) a global phase diagram of our model. Its
core element is the interlayer spin-singlet dimers in the $d^7$ Mott insulator
stabilized by orbital splitting. The superconductivity in the electron doped
spin-singlet phase is our main finding, which is a robust result based on the
textbook treatment of a weakly interacting Fermi liquid coupled to the gapped
singlet-triplet excitations. These excitations play the role of optical
phonons in the BCS theory, and their high energy $\sim 0.3\:\mathrm{eV}$ and
sizable coupling $J_\mathrm{H}$ to conduction spins should give rise to
high-temperature superconductivity. The magnetic order at
$\delta>\delta_\mathrm{crit}$, resulting from competition between RKKY
interactions and spin-singlet gap is also an intrinsic property of the model;
however, its quantitative description is a nontrivial problem left for future
work. From a broader perspective, it is important to extend the model for
small $\Delta$ values, and clarify how its properties, especially
superconductivity, evolve as the \mbox{$3z^2\!-\!r^2$} electrons get
delocalized to form a multiband metal. 

On the materials side, the presence of a large tetragonal field is required to
realize our model. We believe this is exactly the case of
\mbox{La$_3$Ni$_2$O$_7$} films under compressive
strain~\cite{Ko25,Zho25,Liu25,Hao25,Hsu25,Osa25}. Based on Fig.~\ref{fig:PD}(c), 
it is also tempting to speculate that more hole doping of these nominally
$\delta=0.5$ materials may further increase the $T_c$ values.

The recently synthesized oxychloride
\mbox{Sr$_3$Ni$_2$O$_5$Cl$_2$}~\cite{Yam25} and similar mixed-anion compounds
are also promising candidates. No ambient pressure transport data was
reported for \mbox{Sr$_3$Ni$_2$O$_5$Cl$_2$}, but based on a theoretical
prediction of large $\Delta\simeq 1\:\mathrm{eV}$~\cite{Och25}, we expect it
to be $d^7$ Mott insulator similar to \mbox{Sr$_2$NiO$_3X$} 
($X$=Cl, F)~\cite{Tsu22}. Electron doping of this compound may lead to
superconductivity.

To conclude, lifting the orbital degeneracy has profound implications for
the electronic structure and superconductivity in bilayer nickelates. Large
tetragonal field, be it intrinsic or induced by strain, stabilizes the
cupratelike \mbox{$x^2\!-\!y^2$} conduction band, which accommodates all the
pairing strength generated by local spin fluctuations. The key difference from
cuprates, however, is that the spin degrees of freedom in our model reside on
a different (\mbox{$3z^2\!-\!r^2$}) orbital sector, and due to the bilayer
structure, they form local singlets and are thus less harmful for the electron
coherency in the \mbox{$x^2\!-\!y^2$} band -- a great advantage for
superconductivity. This also offers a natural explanation for the recent
observation of a Fermi-liquid behavior well above $T_c$ in
\mbox{La$_2$PrNi$_2$O$_7$} films~\cite{Hsu25}.


{\it Acknowledgments}.~We thank M.~Hepting and B.~Keimer for useful
discussions. G.~Kh. acknowledges support from the European Research Council
under Advanced Grant no.~101141844 (SpecTera). J.~Ch. acknowledges support by
Czech Science Foundation (GA\v{C}R) under Project No.~GA22-28797S and by the
project Quantum Materials for Applications in Sustainable Technologies, Grant
No.~CZ.02.01.01/00/22\_008/0004572.

{\it Data availability}.~The data that support the findings of this
article are openly available~\cite{repository}.


\bibliography{paper}

\begin{thebibliography}{43}
\bibitem [{SKn()}]{SKnote}%
  \BibitemOpen
  \href@noop {} {}\bibinfo {note} {For simplicity, the Slater-Koster
  rules~\cite{Sla54} are used; corrections due to non-cubic symmetry can be
  readily included, but they are not essential here.}\BibitemShut {Stop}%
\end{thebibliography}

\onecolumngrid
\renewcommand\thesection{\Alph{section}}
\renewcommand\thefigure{S\arabic{figure}}


\clearpage
\setcounter{section}{0}
\setcounter{figure}{0}
\setcounter{equation}{0}

\begin{center}
{\large Supplemental Material for}
\vskip 3mm
{\large\bf Orbital Order and Superconductivity in Bilayer Nickelate Compounds}
\vskip 5mm
Giniyat Khaliullin
\vskip 1mm
{\it Max Planck Institute for Solid State Research, \\
Heisenbergstrasse 1, D-70569 Stuttgart, Germany}
\vskip 3mm
Ji\v{r}\'{\i} Chaloupka
\vskip 1mm
{\it Department of Condensed Matter Physics, Faculty of Science, \\
Masaryk University, Kotl\'a\v{r}sk\'a 2, 61137 Brno, Czech Republic}
\end{center}


\vskip 1mm

While the simplified tight-binding (TB) model used in the main text is sufficient for a basic illustration of our model and its underlying physics, a more detailed fitting is needed to capture Fermi surface topology of real materials. In this Supplemental Material we show a possible extension of the TB model for the conduction $c$ electrons of \mbox{$x^2\!-\!y^2$} symmetry, aimed to reproduce the measured Fermi surfaces.

We first recall that virtual hoppings $t_{xz}$ of electrons from \mbox{$x^2\!-\!y^2$}
orbitals to $3z^2\!-\!r^2$ orbital levels, their hybridization with the oxygen $p$ 
as well as other high-energy states generally result in rather diffuse Wannier orbitals,  
as it is well known from e.g. ab-initio studies of cuprates~\cite{Pav01}. The resulting effective bands have the same symmetry as the pure \mbox{$x^2\!-\!y^2$} orbitals, but the corresponding TB model parameters may greatly differ, due to admixture of the other states which provide new hopping processes as demonstrated below. 

In the present case, admixture of the $3z^2\!-\!r^2$ orbitals is most important, as they bridge two NiO$_2$ layers and open the interlayer hopping channel for conduction electrons. Perturbatively, the effective electron operator is a combination
$c_i= \sqrt{1-4\lambda^2}\;x_i + \lambda \;(z_{i+a} + z_{i-a} - z_{i+b} - z_{i-b})$, where $x_i$ corresponds to the central \mbox{$x^2\!-\!y^2$} orbital and the $\lambda$-term adds a superposition of the nearest-neighbor (NN) $3z^2\!-\!r^2$ orbitals.
It has $\mbox{$x^2\!-\!y^2$}$ symmetry as dictated by the bond-dependent signs of the inter-orbital hopping $t_{xz}=\pm \frac{\sqrt3}{4}t$ (see main text). Parameter $\lambda\approx |t_{xz}|/E$ depends on the $e_g$ orbital splitting, $E\sim \Delta$, and using $t\sim 0.6$~eV and $E\sim 1$~eV, we evaluate $\lambda\sim 1/4$ as a rough estimate of the $z$ orbital spectral weight  in the conduction band states.    

The overlap between the renormalized $c_{1,i}$ and $c_{2,i}$ states gives ``vertical'' interlayer hopping $t_\perp=4\lambda^2 t$, where $t$ is $3z^2\!-\!r^2$ orbital overlap along $c$ axis. With $\lambda\sim 1/4$ and $t\sim 0.6$~eV, we estimate $t_\perp\sim 0.15$~eV. Similarly, one also finds finite overlap between $c_1$ and $c_2$ states at longer distances across the bilayer, resulting in the hopping terms $-t_{2\perp}c^\dagger_{1,i}c_{2,i+a+b}$ and $-t_{3\perp}c^\dagger_{1,i}c_{2,i+2a}$ with $t_{2\perp}=-2\lambda^2 t=-t_\perp /2$ and $t_{3\perp}=\lambda^2 t=t_\perp/4$. 

In general, more realistic wave-functions contain also oxygen $p$ states, etc. As an illustration, we consider here a minimal TB model including the interlayer hoppings considered above; this leads to the following dispersions of bonding $\alpha$ and antibonding $\beta$ bands 
\begin{equation}\label{eq:TB}
\varepsilon^{\alpha/\beta}_{\vc k} = 
-4t_1\gamma_{\vc k} -4t_2\gamma'_{\vc k} 
	\,\mp\, (t_\perp 
	         + 4t_{2\perp}\gamma'_{\vc k} + 4t_{3\perp}\gamma''_{\vc k})
\end{equation} 
with the NN form-factor $\gamma_{\vc k}=\frac12(\cos k_x + \cos k_y)$, 
second-NN $\gamma'_{\vc k}=\cos k_x \cos k_y$, and third-NN 
$\gamma''_{\vc k}=\frac12(\cos 2k_x + \cos 2k_y)$. 

\begin{figure*}[h]
\vspace{5mm}

\includegraphics[scale=1.0]{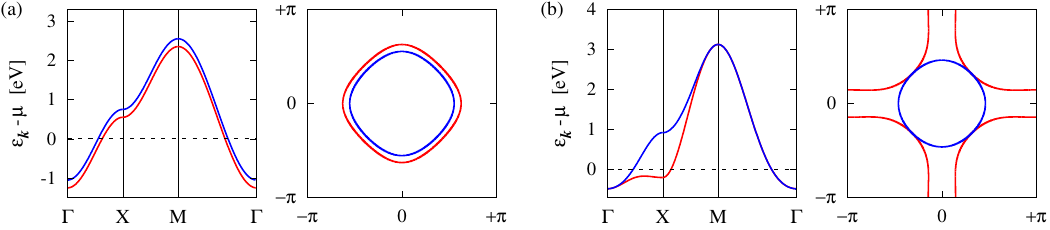}
\caption{Band dispersions and Fermi surfaces for the
tight-binding model \eqref{eq:TB}: 
(a) Basic nearest-neighbor model with 
$t_1=0.45\:\mathrm{eV}$ and $t_\perp=0.1\:\mathrm{eV}$.
(b)~Model including further-neighbor hoppings to reproduce the
observed Fermi surfaces:
$t_1=0.45$,
$t_2=-0.12$,
$t_\perp=0.14$,
$t_{2\perp}=-0.07$, and
$t_{3\perp}=0.035$
(in units of eV).
The band filling corresponds to $0.5$ electron per site.
}\label{fig:bands}
\vspace{-3mm}
\end{figure*}

Figure~\ref{fig:bands} compares the basic TB model and the
extended model (1) with the parameters chosen to fit the measured Fermi surfaces~\cite{Wan25,Sun25}. The main modification of the bands due to the
further-neighbor hoppings takes place around the $X=(\pi,0)$ and $(0,\pi)$
points in the Brillouin zone, leading to a change of the topology of the Fermi surface.

\end{document}